# The Future of High Energy Nuclear Physics in Europe

Jurgen Schukraft, *CERN, 1211 Geneva 23, Switzerland* (schukraft@cern.ch)

*Abstract*— **In less than two years from now, the LHC at CERN will start operating with protons and later with heavy ions in the multi TeV energy range. With its unique physics potential and a strong, state-of-the complement of detectors, the LHC will provide the European, and in fact worldwide Nuclear Physics community, with a forefront facility to study nuclear matter under extreme conditions well into the next decade.**

I. INTRODUCTION

The aim of ultra-relativistic heavy ion physics is the study of the phase diagram of strongly interacting matter under extreme conditions of high temperature and/or high matter density. QCD, the theory of strong interactions, predicts that at sufficiently high energy density there will be a transition from ordinary hadronic matter to a plasma of deconfined quarks and gluons - a transition which took place in the early universe a few microseconds after the Big Bang and which might still play a role today in compact stellar objects.

The field was started in the mid 70's and early 80's at the Bevalac accelerator at Berkeley by a few dozen physicists, mostly from the US, Germany and Japan, with light and later heavy nuclei colliding at an energy of up to 2 GeV/A. When the LHC will come into operation with ion beams in 2008, the available energy in the center of mass system will have increased by four orders of magnitude in 25 years.

Figure 1 shows the center of mass energy available for particle production (i.e. subtracting the rest mass of beam and target nuclei) for different hadron accelerators versus the year of first operation ('Livingston plot'). While the exponential increase in performance is already very impressive for the proton beams used in High Energy Physics, with an energy doubling time of about 4 years, the doubling time for heavy ion beams (1.7 years) is even shorter and rivals Moore's law. Incidentally, the plot also shows that, while LHC as a proton accelerator may be late by a few years, LHC as a heavy ion machine is right on time!

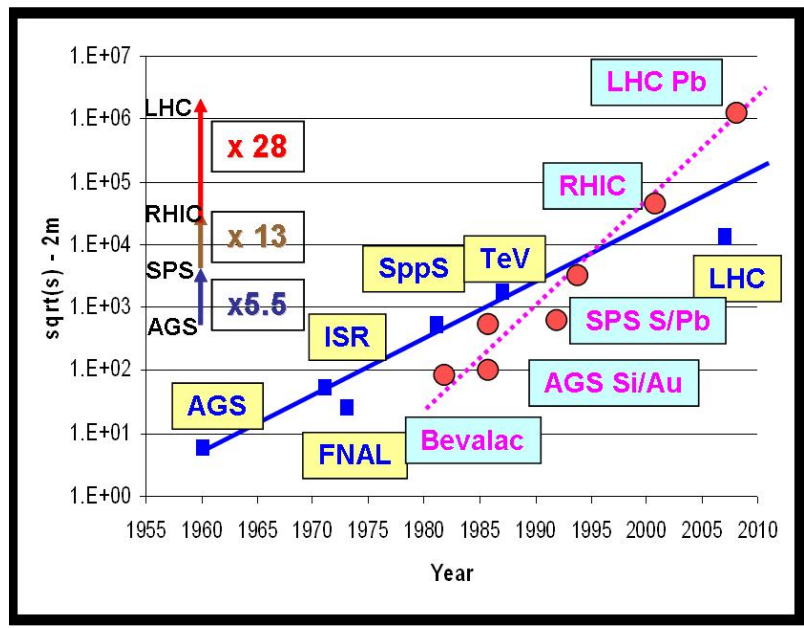

**Figure 1: Total center of mass energy versus time**

This rapid progress was of course only possible by reusing machines, and initially even detectors, built over a longer time scale for Particle Physics; the Relativistic Heavy Ion Collider RHIC remains the only dedicated facility. Today, with more than 2000 physicists active worldwide in this field, ultra-relativistic heavy ion physics has moved in less than a generation from the periphery into a central activity of contemporary Nuclear Physics. Towards the end of this decade, the LHC, with almost thirty times the energy of RHIC, will be at the energy frontier of high energy nuclear physics, and therefore I will concentrate in this contribution on the future of this field in Europe, and indeed worldwide, on the physics prospects with the LHC. The second frontier, at high matter density, to be explored at the future FAIR facility at GSI, will be described elsewhere in these proceedings [1].

## II. HEAVY ION PHYSICS AT THE LHC

*Matter under extreme conditions*

The 'state of matter' at high temperature or density, called the 'Quark-Gluon Plasma' (QGP), is thought to be the ground state of QCD, where partons are deconfined and chiral symmetry is (approximately) restored; i.e. partons are not bound into colourless hadrons and are (approximately) massless. The mission of URHI is threefold: **Search** for the QGP, **measure** its properties, and along the way **discover** QCD in its natural, strongly coupled and non-pertubative regime.

This study of the phases of nuclear matter is relevant also beyond the specific context of QCD, as phase transitions and symmetry breaking are central concepts in the Standard Model of particle physics and the predicted QCD transition is the only one directly accessible to laboratory experiments. It may provide us with some insight into the emerging properties of complex systems involving elementary quantum fields, i.e. how the microscopic laws of physics (the 'QCD equations') give rise to macroscopic phenomena such as phase transitions and critical behavior.

The QGP phase transition, as predicted e.g. by numerical studies on powerful computers ('lattice QCD'), is very well localized at a critical temperature of close to T = 160 MeV ( more than $10^{12}$ Kelvin). However, in practical terms the temperature scale is a very compressed one. More relevant to experiments is the energy density, proportional to the $4^{th}$ root of temperature, which scales (at fixed time) with the number of charged particles created per unit volume, which in turn increases only very slowly - logarithmically - with center of mass energy. Therefore, in order to cross the phase transition, for example from 10% below to 10% above in temperature, we have to increase the energy density by about a factor 2 and the center of mass energy by a factor of 9, i.e. a factor comparable to the difference in energy between the SPS and RHIC machines (see Fig. 1). The big dynamics range in energy which has become available over a short time to experiment is therefore not only a welcome coincidence but a crucial requirement for the exploration of the phase diagram of nuclear matter.

*Physics perspectives at the LHC*

Following the results obtained first at the SPS and since 2000 at RHIC, a lively discussion amongst the experts is ongoing if 'the QGP' has or has not yet been discovered [2, 3]. It is fueled by the fact that some of the results were not widely anticipated, and are very surprising indeed. While many observations are perfectly in line with the telltale signals of the QGP as predicted by QCD, the matter created in these collisions seems to behave more like a very strongly interacting 'perfect liquid', essentially opaque to even very energetic particles, rather then the weakly coupled 'ideal gas' presumed by many. However, while the equations of QCD are perfectly well known, the strength of the 'strong interaction' makes calculations of the properties of dense matter extremely complicated and therefore surprises should not really come as a surprise. I will therefore assume in the following that the QGP, defined pragmatically as 'the stuff at high energy density at which ordinary hadrons are no longer the relevant ingredients', has been discovered prior to the start of the LHC.

So what is left at the LHC to do? The **search** may indeed be over, and the **discovery** phase is well underway with the fantastic results and surprises from RHIC. However, the **measuring** phase has hardly begun. Since heavy ion physics at the LHC was first discussed in 1990, around the time RHIC was approved for construction, it was assumed to be more of a 'precision machine' rather than a 'discovery machine'; not unlike the LEP collider, which came into operation well after the discovery of the vector bosons of the electroweak interaction. It should allow a quantitative and systematic study of the new state of matter, settle a number of still open questions (e.g. the order of the phase transition) and measure, with some precision and in particular with quantifiable errors, relevant parameters, like the ones appearing in its equation of state or the screening mass. Extrapolating from present results, the energy density, size, lifetime and the thermalisation times should all improve significantly and simultaneously, typically by up to an order of magnitude compared to RHIC. In addition, new observables will become accessible at the LHC, in particular hard probes (heavy quarks, jets). The LHC will therefore enter a new regime, very different from the one at the existing machines in both quantitative and qualitative aspects, which should facilitate its task of precision measurements.

With the ongoing flood of results pouring out of RHIC, the tasks for LHC have multiplied and become even more exciting. To start with, we will have to confirm our current interpretation of these results by testing predictions and extrapolations into the LHC regime. The two most prominent examples in this category may be the strong collective flow (the 'perfect fluid') measured at RHIC and the apparent energy independence of $J/\Psi$ suppression. In addition, currently still somewhat inconclusive indications point towards yet another 'new state of matter' dubbed the Colour Glass Condensate, a dense quantum state which may characterize the initial nuclear wave function. If present already at RHIC (in a limited corner of the phase space), it would dominate the initial conditions at LHC and allow the study of QCD in the classical limit of quantum field theory. So also the discovery phase is likely to continue, at both RHIC and LHC, and further surprises should lay ahead.

Finally, not even the search may be completely over yet. In case sufficiently high temperatures can be reached at LHC, the strongly coupled matter (the 'sQGP') may give way to a weakly coupled Quark-Gluon Plasma (presumably to be called the 'iQGP'), governed by the Stefan-Boltzmann laws of an ideal parton gas.

*Signals and Observables*

Conventional wisdom predicts that 'soft' (i.e. low momentum) physics signals, while necessary to measure in order to constrain models, will essentially follow smooth extrapolations from lower energies and 'hard' physics signals will be the focus of interest at LHC. However, with almost 30 times the center-of-mass energy of RHIC, LHC will be a huge step forward and 'conventional wisdom' has not been a very reliable guide in this field even in the past. While some expectations turned out to be correct at RHIC (e.g. particle ratios, jet quenching), others have been wrong (particle multiplicity, $J/\Psi$ suppression) and some of the most important insights have come from completely unexpected corners (elliptic flow, baryon anomaly) [3]. The experimental program at LHC should therefore be open minded, with some emphasis on hard probes unique to this machine but sufficiently broad to cover the complete spectrum of observables.

Hard parton scattering processes are extremely useful tools in heavy ion reactions. Their cross section can be calculated fairly precisely, therefore allowing precision measurements, and they take place at the earliest stages of the collisions and therefore can probe the matter in the initial, high temperature phase. While very rare at the SPS energy, they are of roughly equal importance to soft processes at RHIC and should dominate the event properties at LHC. Rates at the current statistical limit of RHIC, around 20 GeV, will be larger by more than three orders of magnitude and the kinematic reach will extend well above 200 GeV. Heavy quarks (charm and beauty) will be abundantly produced and with up to 200 heavy quark pairs per central event, charm quarks will play a similar role at LHC as the much lighter strange quarks did at the SPS.

The melting of heavy bound quarkonia states is a prime signal of deconfinement and at LHC, the full spectrum of these states (J/Ψ, Ψ', Υ, Υ', Υ'') will be available for analysis. The recent, very surprising if still preliminary, observation that J/Ψ suppression at RHIC is rather comparable to the one observed at the SPS, has added significant interest to these measurements at LHC. If the effect observed at SPS and RHIC are both due to the melting of higher charm resonances (Ψ', χ), as would be consistent with more recent lattice QCD calculations, the J/Ψ itself may melt only at the higher temperatures reached at the LHC. In case the similarity between SPS and RHIC results are due to a fortuitous cancellation of suppression and regeneration processes, the latter should become much stronger at LHC and we should observe a very strong J/Ψ enhancement. While the measurement of J/Ψ production at the LHC was considered up to very recently of marginal interest at best, it has become now crucial to unravel and interpret the results from lower energy.

### III. THE LHC HEAVY ION PROGRAM

*LHC machine*

The LHC as a high energy proton machine was first discussed in the early '80's, and the LEP tunnel was made sufficiently big to eventually accommodate a large hadron collider. Its use for heavy ion physics was first mentioned around 1990, to the best of my knowledge by Carlo Rubbia, then DG of CERN. The project was approved, in two stages, in 1994 and 1996, and first collisions are expected late in 2007.

The LHC will be able to accelerate protons or ions to a maximum center of mass energy of 14 TeV x Z/A (e.g. 14 TeV for pp and 5.5 TeV/nucleon for Pb-Pb) and provide both collisions for symmetric systems (pp, AA) as well as proton-nucleus collisions (pA). Like the SPS, it will operate for about four weeks per year with ions and the rest of the time with proton beams. The nominal luminosity is $10^{34}$ cm$^{-2}$s$^{-1}$ for pp, $10^{27}$ cm$^{-2}$s$^{-1}$ for Pb-Pb and $> 10^{30}$ cm$^{-2}$s$^{-1}$ for lighter ions.

*LHC experiments*

Four large experiments are being constructed for the LHC: Atlas (Fig. 2) and CMS (Fig. 3) are general purpose high energy physics detectors whose main aim is to search for the Higgs boson and physics beyond the standard model; LHCb is a detector focused on CP violation studies with B quarks; ALICE (Fig. 4) is a general purpose experiment dedicated to heavy ion physics. While both Atlas and CMS intend to participate in the heavy ion program, they will concentrate largely on a subset of hard processes which can be measured in these detectors.

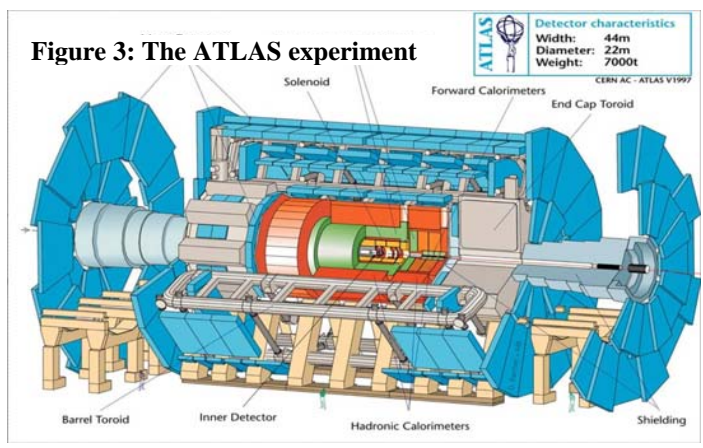

**Figure 3: The ATLAS experiment**

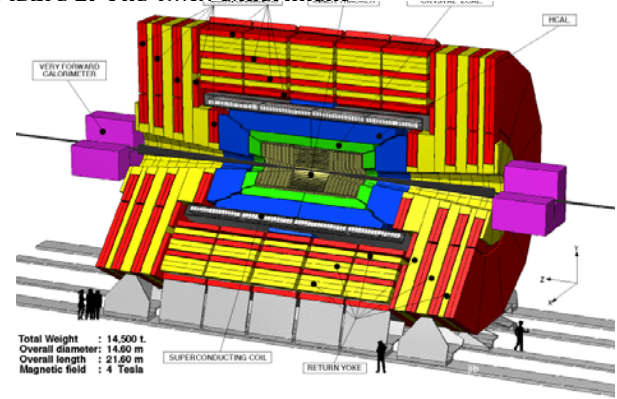

**Figure 2: The CMS experiment**

The dedicated ALICE experiment, which currently includes almost 1000 physicists from over 80 institutions in some 30 countries, will cover essentially all observables thought to be relevant; from very low momenta where the bulk of the particles are produced, up to several hundred GeV, far into the region of hard processes. With a weight of 10,000 tons, a size of 16 x 26 meters, and almost two dozen different detector systems, it is the largest and most complex experiment ever constructed in Nuclear Physics. Special emphasis has been put on a robust tracking system to cope with the very large number of particles expected at LHC (up to 10,000 in the acceptance of the experiment) and it employs essentially all known techniques for particle identification. It includes a number of novel detector technologies specifically developed for use in heavy ion physics in an intense R&D period lasting several years. Several of these technologies are now also being deployed for the upgrade of some of the RHIC experiments.

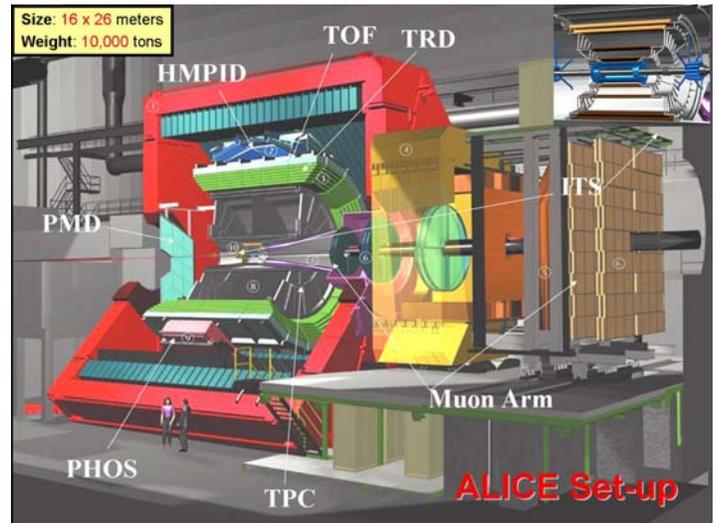

Figure 4: The ALICE experiment

*LHC Heavy Ion program*

Like at the SPS and at RHIC, the initial heavy ion program will take of order 10 years to cover a minimum of different run configurations. It will include several years of running at the highest energy and with the largest nuclei (Pb), p-nucleus running to measure cold nuclear matter effects (and possible study the Colour Glass Condensate in a simpler system), one or two light ion species and a limited number of energy scans in order to look for systematic trends and/or threshold behavior. During this time, some modest detector upgrades are likely, as required by early results and physics priorities.

The long term prospects for heavy ion physics at the LHC are very difficult to predict and will become clear only after first physics results will be available. An increase in beam energy would seem unlikely and of limited benefit. A big factor would in any case be required to increase significantly the energy density (which grows only logarithmically with energy). An increase in luminosity however is quite possible and desirable, as a number of signals are statistically limited even at the LHC (e.g. ϒ production, γ-jet correlations). A modest increase of order five may be feasible, but by no means trivial, as several factors currently limit the machine performance for heavy ions. Such a luminosity upgrade may come around 2015 in parallel with the upgrade in proton luminosity ('SuperLHC') currently under discussion.

Beyond the LHC, a second front will open up in the study of nuclear matter at high density, as opposed to high temperature, at the FAIR facility approved at GSI and, potentially, with electron-nucleus scattering at a facility like eRHIC (or eLHC?), where dense partonic systems can be probed inside cold nuclei.

IV. FUTURE OF THE SPS

The SPS heavy ion program, which started in 1986, came to an end in 2003. However, the relevant CERN committee has examined in 2004 the physics case for a continued fixed target program at CERN in the energy range from 20 to 200 GeV/nucleon. It concluded that the SPS energy range is still unique and interesting for two reasons: First, there remains some 'unfinished business', in particular a high statistics and precision measurement of the low mass lepton pair continuum which is related to the chiral symmetry restoration expected in the QGP. The third generation experiment NA60 could only take data in a single ion run with Indium beams, resulting in a spectacular and intriguing signal. Secondly, 'new physics' may still be hiding, if indeed the QCD 'critical endpoint' where located in the SPS energy range. On this topic, however, the committee felt that further progress on both the theoretical and experimental front would be required before a new program could be launched. Having recognized a potential physics case, it is quite possible that a new heavy ion fixed target program will run at CERN in parallel to the LHC, i.e. after 2009. It remains to be seen, however, if sufficient human and financial resources can be found for such an effort, in parallel with the other programs at RHIC, LHC and FAIR.

V. CONCLUSION

In the incredibly short time span of little over 25 years, the study of the phases of nuclear matter will have evolved from light ion reactions at a laboratory energy of one GeV/nucleon to using heavy projectiles at a center-of-mass energy of several TeV/nucleon. Within less than a generation, the field has moved from the periphery to become a mainstream activity of contemporary Nuclear Physics. The LHC, the latest and quite possibly last, nuclear collider at the high energy frontier will open for business in 2007, providing a unique and very different opportunity to study strongly interacting matter. With state of the art detectors and a large and enthusiastic, worldwide community, we can look forward to exploring the 'wonderland' that may open up in the years to come.

VI. REFERENCES

[1] 'Nuclei and Hadrons: Perspective for Europe', W, Henning, these proceedings  
[2] U. Heinz and M. Jacob, preprint nucl-th/0002042  
[3] 'First three years of operation of RHIC', Nucl. Phys A757 (2005) issues 1-2